
\magnification=1200
%
%
\def\half{{1\over 2}}
\def\gmu{{\gamma_{\alpha\beta}^\mu}}
\def\a{{\alpha}}
\def\b{{\beta}}
\def\g{{\gamma}}
\def\d{{\delta}}
\def\e{{\epsilon}}
\def\dev{{\delta^S_{[\e^\a,v_j^\a]}}}
\def\devh{{\delta^S_{[\hat\e^\a,\hat v_j^\a]}}}
\def\devo{{\delta^S_{[\half(\g^{\mu\nu}\e)^\a,\half(\g^{\mu\nu}v_j)^\a]}}}
\tolerance=5000
\footline={\ifnum\pageno>1
       \hfil {\rm \folio} \hfil
    \else \hfil \fi}

\overfullrule=0pt 
\baselineskip=18pt
\raggedbottom
\centerline{\bf A Ten-Dimensional Super-Yang-Mills Action}
\centerline{\bf with Off-Shell Supersymmetry}
\vskip 12pt
\centerline{Nathan Berkovits}
\vskip 12pt
\centerline{Maths Dept., King's College, Strand, London, WC2R 2LS, United
Kingdom}
\centerline{e-mail: udah101@oak.cc.kcl.ac.uk}
\vskip 12pt
\centerline {KCL-TH-93-11}
\vskip 12pt
\centerline {August 1993}
\vskip 24pt
\centerline {\bf Abstract}
After adding seven auxiliary scalar fields, the action for ten-dimensional
super-Yang-Mills contains an equal number of bosonic and fermionic non-gauge
fields. Besides being manifestly Lorentz and gauge-invariant, this action
contains nine spacetime supersymmetries whose algebra closes off-shell.
Octonions provide a convenient notation for displaying these symmetries.
\noindent
\vfil\eject
\vskip 12pt
Ten-dimensional super-Yang-Mills is of special interest to physicists
both because it contains the maximum number of supersymmetries consistent
with Yang-Mills and because it is described by the massless sector of the
ten-dimensional superstring. At the present time, there are two different
actions used for describing this theory.

The first action is constructed from a bosonic vector and a fermionic spinor
and is both manifestly Lorentz-invariant and gauge-invariant.$^1$
However because a vector and a spinor contain different numbers of non-gauge
degrees of freedom in ten dimensions, the spacetime supersymmetries of
this action only close on-shell. In other words, the commutator of two
supersymmetry transformations is not just a spacetime translation plus
a gauge transformation, but includes other contributions which only
vanish after using equations of motion.

The other action used to describe ten-dimensional super-Yang-Mills is
light-cone gauge-fixed and therefore contains only the eight physical
bosonic and fermionic fields.$^2$
Although eight of the spacetime supersymmetries
now close off-shell, this action is non-local in $x^-$
and neither Lorentz invariance nor gauge invariance is manifest.

In this paper, a new action for ten-dimensional super-Yang-Mills is
constructed which combines advantages of these two earlier versions.
In addition to being manifestly Lorentz and gauge-invariant, it contains
nine spacetime supersymmetries which close off-shell. This action is:
$$S=\int d^{10}x\, Tr(-\half F^{\mu\nu}F_{\mu\nu} +i
 \psi^\a \gmu D_\mu \psi^\b
- \sum_{j=1}^7 G_j G_j ) \eqno(1)$$
where $D_\mu=\partial_\mu +g A_\mu$, $F_{\mu\nu}=[D_\mu ,D_\nu ]$,
$A_\mu$ is the usual vector gauge field, $\psi^\a$ is the usual
sixteen-component Majorana-Weyl spinor, $G_j$ is a bosonic auxiliary scalar
for $j=1$ to 7, and all Lie-algebra indices have been suppressed.
The above action is invariant under the following supersymmetry
transformation:
$$\dev A^\mu =i\e \g^\mu \psi ,\quad
\dev \psi^\a=-\half(\e \g^{\mu\nu})^\a F_{\mu\nu}+
\sum_{j=1}^7 v_j^\a G_j ,$$
$$ \dev G_j =-iv_j \g^\mu D_\mu \psi ,\eqno(2)$$
where $\e^\a$ and $v_j^\a$ are eight bosonic Majorana-Weyl
spinors satisfying
$$v_j \g^\mu v_k -\d_{jk}\, \e\g^\mu \e= v_j \g^\mu \e =0 \eqno(3)$$
for $j,k=1$ to 7 and $\mu=0$ to 9 (for a given non-zero $\e^\a$,
the $v_j^\a$'s are determined up to an SO(7) rotation).
Note that $\dev$ is fermionic and corresponds to a single supersymmetry
generator without the anti-commuting parameter (if desired, $\dev$ can be
made bosonic by including the anti-commuting parameter $\kappa$, in which case
the commutator becomes proportional to $\kappa_1 \kappa_2$).

Using the property that $\half(\e\g^\mu\e)\g_\mu^{\a\b}=\e^\a\e^\b
+\sum_{j=1}^7 v_j^\a v_j^\b$ and ten-dimensional $\g$-matrix identities,
it is easy to check that
$$\dev\dev A_\nu=-i F_{\mu\nu}(\e\g^\mu\e),\quad
\dev\dev \psi^\a=-iD_\mu\psi^\a (\e\g^\mu\e),$$
$$\dev\dev G_j=-iD_\mu G_j (\e\g^\mu\e).\eqno(4)$$
Therefore, for a given $\e^\a$ and $v_j^\a$, the anti-commutator of
$\dev$ with itself is proportional to a spacetime translation
$\d^{P^\mu}_{\e\g^\mu\e}$
(i.e., $x^\mu\to x^\mu+\e\g^\mu\e$) plus a gauge transformation
$\d^\Lambda_{A_\rho(\e\g^\rho\e)}$ (i.e., $A_\mu\to A_\mu+ D_\mu A_\rho
(\e\g^\rho\e)$). Furthermore, since the commutator of the
Lorentz generator $M_{\mu\nu}$ with $\dev$ is equal to $\devo$,
$$\{ \dev ,\devo \}=[M_{\mu\nu} ,\dev \dev]=
[M_{\mu\nu}, -i(\d^{P^\rho}_{\e\g^\rho\e}
+\d^\Lambda_{A_\rho(\e\g^\rho\e)}) ]$$
$$=-i( \d^{P^\rho}_{\e\g^\rho\g^{\mu\nu}\e}+
\d^\Lambda_{A_\rho(\e\g^\rho\g^{\mu\nu}\e)}).\eqno(5)$$
So the anti-commutator of $\dev$ with $\devo$ is also proportional to a
spacetime translation plus a gauge transformation.

However if $[\hat\e^\a, \hat v_j^\a]$ does not equal $[(\g^{\mu\nu}
\e)^\a, (\g^{\mu\nu} v_j)^\a] f_{\mu\nu}$
for some $f_{\mu\nu}$,
$\{ \dev ,\devh \}$ is not just a translation plus a gauge transformation,
but includes other contributions which only vanish on-shell. The maximum
number of independent spacetime supersymmetries which closes off-shell is
therefore nine, since for more than nine, there always exists at least
two supersymmetries, $\dev$ and $\devh$, which do not satisfy
$[\hat\e^\a, \hat v_j^\a]=[(\g^{\mu\nu}\e)^\a,
(\g^{\mu\nu} v_j)^\a] f_{\mu\nu}$ for any $f_{\mu\nu}$.
An example of nine independent
supersymmetries which do form a closed algebra off-shell
is $\dev$ and $\d^S_{[(\g^{+\mu }\e)^\a, (\g^{+\mu }v_j)^\a]}$
for $\mu=1$ to 8, where $(\g^-\e)_\a=0$ and
$\g^\pm_{\a\b}=\g^0_{\a\b}
\pm\g^9_{\a\b}$.

The action of equation (1) is of course analogous to off-shell
supersymmetric actions for super-Yang-Mills in three, four, or six
dimensions, in which zero, one, or three bosonic auxiliary scalars
are added to the bosonic vector and fermionic spinor. Just as the
off-shell closure of supersymmetry transformations in these dimensions
is related to the real, complex, and quaternionic division algebras,$^3$
the off-shell closure of the supersymmetry transformations in equation (2)
can be related to the octonionic division algebra.

It is well-known that the SO(9,1) $\g$-matrices can be represented
as $2\times 2$ octonion-valued hermitian matrices, $\tilde
\g^\mu_{c\dot d}$, where
$\tilde\g^0_{c\dot d}=\sigma^0_{c\dot d}$, $\tilde\g^j_{c\dot d}=
(i\sigma^2_{c\dot d}) e_j$ for $j=1$ to 7,
$\tilde\g^8_{c\dot d}=\sigma^1_{c \dot d}$, $\tilde
\g^9_{c\dot d}=\sigma^3_{c\dot d}$, $\sigma^r_{c \dot d}$ are the usual
$2\times 2$ Pauli-matrices (indices are raised and lowered using the
epsilon tensor),
and $e_j$ for $j=1$ to 7 are the imaginary octonions ($e_j=-\bar e_j$
and $e_j e_k +e_k e_j=-2\d_{jk}$ for $j,k=1$ to 7).
In this representation, an SO(9,1)
Majorana-Weyl spinor, $\psi^\a$, can be written as two octonionic
components, $\tilde\psi_1=\sum_{a=1}^8 e_a (\g^+\psi)_a$ and
$\tilde\psi_2=\sum_{a=1}^8 e_a (\g^-\psi)_{a+8}$, where $e_8=1$.$^4$

If the spacetime supersymmetry parameter, $\e^\a$, satisfies
$\tilde\e_2=\bar{\tilde\e}_2$, then the $v_j^\a$'s can be
chosen such that $\tilde v^c_j =e_j\tilde \e^c $ for $c=1$ to 2 and
$j=1$ to 7 since
$$\tilde v^c_j(\tilde\g^\mu_{c\dot d}\bar{\tilde v}^{\dot d}_k)
+(\tilde v^c_k\tilde\g^\mu_{c\dot d})\bar{\tilde v}^{\dot d}_j=
(e_j\tilde\e^c )(\tilde\g^\mu_{c\dot d}(\bar{\tilde \e}^{\dot d}\bar e_k))+
((e_k\tilde\e^c )\tilde\g^\mu_{c\dot d})(\bar{\tilde \e}^{\dot d}\bar e_j)
=\eqno(6)$$
$$
\tilde\g^\mu_{{c\dot d}}((\bar{\tilde \e}^{\dot d}\bar e_k)(e_j\tilde\e^c))
+\tilde\g^\mu_{{c\dot d}}((\bar{\tilde \e}^{\dot d}\bar e_j)(e_k\tilde\e^c))
=2\d_{jk} \tilde\g^\mu_{c\dot d}
(\bar{\tilde \e}^{\dot d}\tilde\e^c)=
2\d_{jk} \tilde\e^c(\tilde\g^\mu_{c\dot d}\bar{\tilde\e}^{\dot d}),$$
and therefore equation (3) is satisfied. The restriction on $\tilde\e_2$
comes from the non-associativity of octonions and reduces the number of
independent $\e^\a$'s from sixteen to nine.

It is straightforward to show that the spacetime supersymmetry transformations
of equation (2) form a closed algebra off-shell when parameterized by
these nine $\e^\a$'s, and can be expressed in octonionic notation as:
$$\d A^\mu ={i\over 2}(~\tilde \e (\tilde\g^\mu\bar{\tilde\psi})+
(\tilde \psi \tilde\g^\mu ) \bar{\tilde\e}~),\quad
\d \tilde\psi_c=-\half((\tilde \e \tilde\g^\mu)
\tilde\g^{\nu})_c
F_{\mu\nu}+\tilde\e_c G,$$
$$\d G={i\over 2}(~{\tilde\e}(\tilde \g^\mu D_\mu\bar{ \tilde\psi})
-(D_\mu {\tilde\psi}\tilde\g^\mu)\bar{\tilde\e}~) ,\eqno(7)$$
where $G=\sum_{j=1}^7 G_j e_j$ and contracted SU(2) spinor indices
have been suppressed.

If the octonions in equation (7) are replaced by real, complex, or
quaternionic numbers, one obtains the supersymmetry transformations
for super-Yang-Mills in three, four, or six dimensions.$^3$ Because of
associativity of the division algebra, $\tilde\e_2$ is unrestricted in these
dimensions.

There are various possible applications of this new ten-dimensional
super-Yang-Mills action. For example, one could try to add more gauge
fields so that the anticommutator of two supersymmetry transformations
is a pure spacetime translation. This might help in making the off-shell
supersymmetry manifest since it would allow non-Wess-Zumino gauge choices.
Another possible application would be to construct a new action for
ten-dimensional supergravity such that some of the spacetime
super-reparameterizations close off-shell in a similar manner.

It is intriguing that division algebras also play a useful role in
understanding the symmetries of the Green-Schwarz superstring,$^5$ which
describes super-Yang-Mills in first-quantized language. Since
string field theory translates first-quantized into second-quantized
language, it is natural to conjecture that string field theory for
the Green-Schwarz superstring somehow relates the two roles played
by the division algebra.
\vskip 24pt
\centerline {\bf Acknowledgements}
\vskip 12pt
I would like to thank A. Galperin, P. Howe, M. Rocek, J. Schray,
W. Siegel, E. Sokatchev, P.Townsend and J. Yamron for useful discussions.
This work was supported by an SERC grant.

\vskip 24pt

\centerline{\bf References}
\vskip 12pt

\item{(1)} Brink,L., Schwarz,J.H., and Scherk,J., Nucl.Phys.B121 (1977), p.77.

\item{(2)} Brink,L., Lindgren,O., and Nilsson,B.E.W., Nucl.Phys.B212 (1983),
p.401.

\item{(3)} Kugo,T. and Townsend,P., Nucl.Phys.B221 (1983), p.357.
(1986), p.93.

\item{(4)} Chung,K.-W. and Sudbery,A., Phys.Lett.B198 (1987), p.161.

\item{} Ramond,P., ``Introduction to Exceptional Lie Groups and Algebras'',
Caltech preprint CALT-68-577 (1976).

\item{(5)} Fairlie,D.B. and Manogue,C.A., Phys.Rev.D36 (1987), p.45.

\item{}  Bengtsson,I. and Cederwall,M., Nucl.Phys.B302 (1988), p.81.

\item{}  Delduc,F., Galperin,A., Howe,P., and Sokatchev,E.,
``A twistor formulation of the heterotic D=10 superstring with
manifest (8,0) worldsheet supersymmetry'', preprint BONN-HE-92-19,
JHU-TIPAC-920018, ENSLAPP-L-392-92, July 1992.

\item{}   Berkovits, N., Phys.Lett.B241 (1990), p.497.

\end